# scientific reports

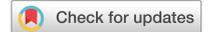

OPEN

# Beyond 5 GHz excitation of a ZnO-based high-overtone bulk acoustic resonator on SiC substrate

Padmalochan Panda, Soumyadip Chatterjee, Siddharth Tallur & Apurba Laha✉

This work reports on the fabrication and characterization of an Au/ZnO/Pt-based high-overtone bulk acoustic resonator (HBAR) on SiC substrates. We evaluate its microwave characteristics comparing with Si substrates for micro-electromechanical applications. Dielectric magnetron sputtering and an electron beam evaporator are employed to develop highly *c*-axis-oriented ZnO films and metal electrodes. The crystal structure and surface morphology of post-growth layers are characterized using X-ray diffraction, atomic force microscopy, and scanning electron microscopy techniques. HBAR on SiC substrate results in multiple longitudinal bulk acoustic wave resonances up to 7 GHz, with the strongest excited resonances emerging at 5.25 GHz. The value of *f.Q* (Resonance frequency.Quality factor) parameter obtained using a novel *Q* approach method for HBAR on SiC substrate is 4.1 × 10$^{13}$ Hz, which to the best of our knowledge, is the highest among all reported values for specified ZnO-based devices.

High-performance bulk acoustic wave (BAW) resonators have received a great deal of attention over the past few decades due to their potential use as radio frequency (RF) sources, sensors, filters, and actuators.[1,2] Quartz crystal resonator (QCR), which typically operates in the several MHz to tens of MHz range, is a common type of BAW resonator. Another type of BAW resonator is the High Overtone Bulk Acoustic Resonator (HBAR), also referred to as a composite resonator composed of a piezoelectric layer sandwiched between two metal electrodes on a low acoustic loss substrate.[3,4] With a simple yet robust structure, a compact size, and an impressively high-quality factor (*Q*), HBAR has the ability to demonstrate highly acute resonances (*f*) at GHz frequencies and above than that of QCR. Owing to these properties, HBAR has emerged as a viable contender for use in the low-noise oscillators, sensors, and phonon sources in quantum acousto-dynamics systems.[5–8] Recently, there has been a noticeable growth in interest in the development of highly sensitive smart physical, chemical, and biological sensors based on acoustic resonators for noninvasive detection in real-time applications without utilizing any external reagents/chemicals. The working principle here is to integrate a biological/chemical element with the physical transducer of the acoustic device since it is sensitive to the atomic, ionic, or molecular chemical bond strength at the microwave frequency range.[9,10] Hence, HBAR can be widely employed to analyze a broad range of small volumes of fluidic materials, including human physiological fluids, and is suitable for Lab-on-a-Chip (LoC) systems.[11–14]

Usually, lead zirconate titanate (PZT), gallium nitride (GaN), aluminium nitride (AlN), and zinc oxide (ZnO) films have undergone exhaustive research for acoustic devices.[15–18] PZT offers a distinctive variety of features, including a very high piezoelectric constant and effective electromechanical coupling ($k_{eff}^2$) value. However, it is not suited for HBAR applications due to its lower acoustic velocities, higher acoustic wave attenuation, and challenges in preparing thin films.[1,11] GaN films are substantially less prevalent due to their poor piezoelectric properties and low $k_{eff}^2$ value.[19] Although AlN films possess high acoustic velocity compared to ZnO films, it again suffers from a low $k_{eff}^2$ value.[19] Among the numerous piezoelectric materials described above, ZnO films with enhanced electro-acoustic characteristics have been found to be the most promising for the development of HBAR devices. Nevertheless, as reported in the literature, HBAR devices with ZnO piezoelectric layer have mostly been restricted to sapphire substrates with a *f.Q* product value of around 4.8 × 10$^{13}$ Hz using Lakin's *Q* method.[4,20,21] Also, ZnO-based HBAR has previously been demonstrated on quartz and diamond substrate; however, they exhibit a lower *f.Q* value of about 1.1 and 0.2 × 10$^{13}$ Hz, respectively.[4,22] Apart from the above substrate, silicon carbide (SiC) is also known as the low acoustic-loss (0.4 dB/cm @ 1 GHz) and high acoustic-velocity substrate when compared to sapphire and diamond substrates making it suitable for HBAR devices and conveniently compatible with surface micromachining processes.[8,22–24] Furthermore, SiC is frequently used in

Electrical Engineering Department, Indian Institute of Technology Bombay, Mumbai 400076, India. ✉email: laha@ee.iitb.ac.in

high-temperature and high-power electronic devices due to its high hardness, high thermal conductivity, chemical resistance, and so on. SiC substrates also play a vital role in the new generation of hybrid quantum sensors and systems since they generate high stress at GHz frequencies than other substrates do.[22] It is, therefore, imperative and pertinent to investigate the microwave resonant properties of ZnO-based HBAR on SiC substrates.

In the present work, we report a novel device comprising a *c*-axis oriented ZnO piezoelectric film deposited on a Pt/Ti-coated SiC substrate to realize an efficient yet simple and scalable heterostructure for fabricating high-overtone bulk acoustic resonators. The detailed HBAR characteristics of the ZnO film on a SiC substrate are compared to those of a Si substrate over a broad frequency range. HBAR on SiC performs well with multiple longitudinal bulk acoustic wave resonances up to 7 GHz and a *f.Q* value up to $4.1 \times 10^{13}$ Hz, which is superior to any reported *f.Q* value among the specified ZnO-based HBAR on any other substrate.

## Results and discussion

**Crystal structure and morphology of ZnO films.** A $650 \pm 20$ nm thick ZnO piezoelectric film was grown on the Pt/Ti-coated Si and SiC substrate using the RF sputtering method. The structural properties of the synthesized ZnO film on the Pt/Ti coated oxidized Si and SiC substrates were investigated using high-resolution X-ray diffraction (HRXRD, M/s. Rigaku, Japan), and the results are shown in Fig. 1. ZnO layers deposited on Pt/Ti/SiC exhibit a stronger (0002) orientation as compared to ZnO deposited on an oxidized Si substrate with a Pt/Ti coating. The (0002) rocking curve for ZnO on SiC is depicted in the inset of Fig. 1, with a full width at half maximum (FWHM) of 2.45°. This outcome is consistent with the cross-sectional field-emission scanning electron microscope (SEM, M/s. Carl Zeiss, Germany) observation at 3 kV operating voltages, as shown in Fig. 2a,b. On the SiC substrate, ZnO is shown to have a much better columnar microstructure normal to the substrate surface than on the Si substrate. The surface morphology was measured using atomic force microscopy (AFM, Asylum Research, M/s. Oxford Instruments, UK) with a non-contact cantilever single-crystal silicon tip of size 10 nm in the tapping mode, and a representative result for ZnO/Pt/Ti/SiC is displayed in Fig. 2c. The ZnO films exhibit RMS surface roughness of 9.7 ± 0.3 nm and 4.9 ± 0.2 nm for ZnO/Pt/Ti/Si and ZnO/Pt/Ti/SiC, respectively.

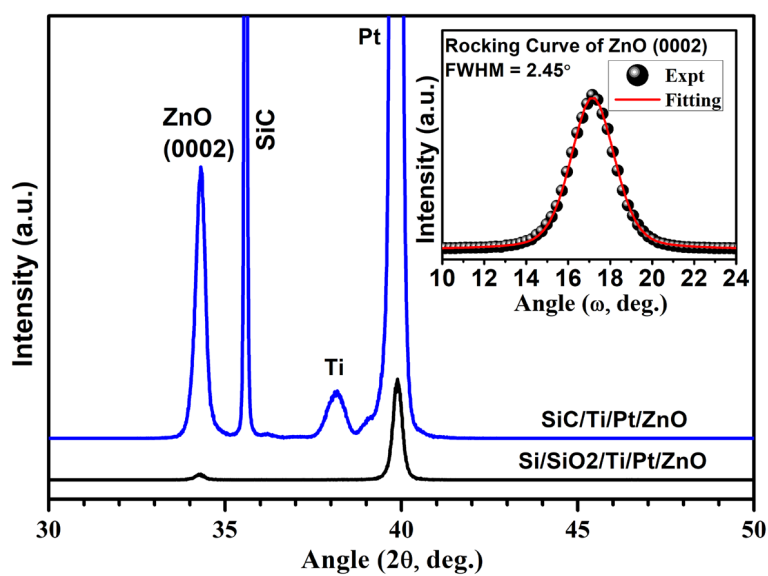

**Figure 1.** The XRD profile of ZnO films on Pt/Ti/*SiO₂*/Si and Pt/Ti/SiC substrate and (inset) rocking curve of (0002) peak of ZnO film on Pt/Ti/SiC substrate.

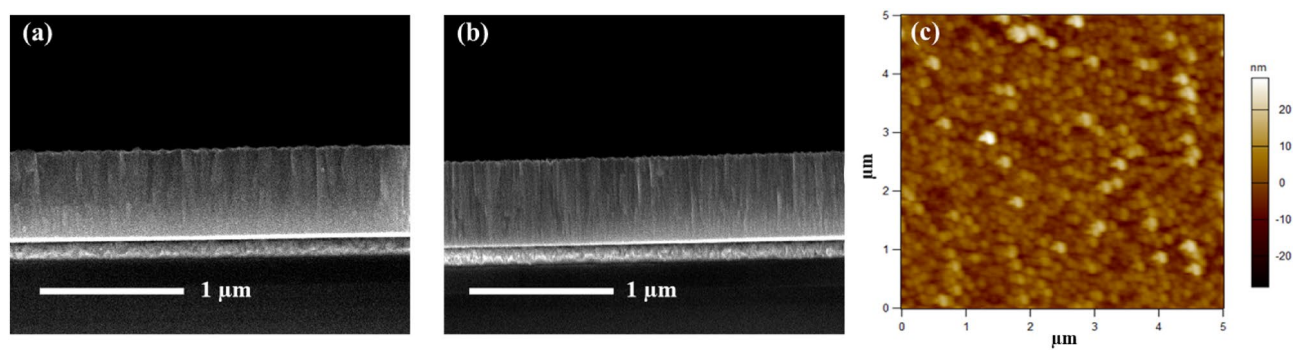

**Figure 2.** SEM cross-sectional micrograph of ZnO film grown on (**a**) Pt/Ti/Si substrate, (**b**) Pt/Ti/SiC substrate, and (**c**) AFM image of ZnO film grown on Pt/Ti/SiC substrate.

**Electro-acoustic characterization.** The measured reflection coefficient ($S_{11}$) parameter of the fabricated ZnO-based HBAR on the Si and SiC demonstrates multiple resonances over a very wide band (up to 7 GHz), as shown in Fig. 3a,b, respectively. The strongest excited resonances for the ZnO-HBAR on Si and SiC are centered at frequencies of 1.85 GHz and 5.25 GHz, respectively. ZnO has a longitudinal acoustic velocity of approximately 6400 m/s and a shear acoustic velocity of about 2770 m/s.[25] The strong resonance frequency ($f_n$) of the HBAR can be approximated using the expression $f_n = v_a/2t$ where $v_a$ and t are the acoustic velocity and thickness of the piezoelectric film, respectively.[4,26] In literature, it is reported that if the c-axis of ZnO film is perfectly oriented with respect to the normal of the substrate surface (Zero tilted angle), then the effective electromechanical coupling coefficient for longitudinal acoustic propagation ($k^2_{L,eff}$) is around 8.53% and for shear acoustic propagation ($k^2_{S,eff}$) is 0%. However, if the c-axis of the ZnO is tilted at any angle with the normal of the substrate, then the $k^2_{S,eff}$ for shear mode acoustic wave takes precedence over the $k^2_{L,eff}$ for longitudinal mode acoustic wave.[25] In this study, the c-plane of ZnO film is not highly oriented along the normal of the Si substrate, as evidenced by the XRD study and cross-sectional SEM micrographs in Figs. 1 and 2. Additionally, Si has relatively high longitudinal acoustic propagation losses (8.3 dB/cm @ 1 GHz), which are contrasted with its shear acoustic propagation losses (3.0 dB/cm @ 1 GHz) and a lower acoustic velocity than the other substrate materials.[23] Hence, HBAR on a Si substrate only exhibits shear resonance. On the other hand, ZnO film is highly oriented along the (0002) direction on the SiC substrate, which is observed from the rocking curve analysis of XRD with an FWHM of 2.45° and cross-sectional SEM micrograph in Fig. 2 compared to a Si substrate. Additionally, SiC is widely known for having low acoustic losses in both longitudinal and shear acoustic propagations (0.4 and 0.3 dB/cm @ 1 GHz) and a high acoustic velocity when compared to Si substrates.[23] Therefore, the ZnO-based HBAR mounted to a SiC substrate demonstrates both shear and longitudinal resonance.

Figure 3c represents the measured impedance, or $Z_{11}$ parameter, of the HBAR close to the strongest excited resonances on the SiC. The frequency range between each narrow resonance depends on the thickness ($t_s$) of the substrate since the acoustic energy from the piezoelectric layer is coupled to it. This frequency spacing ($\Delta f_{overtone}$) between narrow resonances is determined as $\Delta f_{overtone} = v_s/2t_s$, where $v_s$ is the acoustic velocity of the substrate.[4] The computed acoustic velocity from the equation is often a few percent lower than the real acoustic velocity because this expression is produced by ignoring the action of the piezoelectric layer on the substrate. The equation below describes the discrepancy between the calculated and actual acoustic velocities.[4]

$$v_{actual} - v_{calc} = v_{calc}\frac{\rho_p l_p}{\rho_s l_s} \qquad (1)$$

where the mass density and thickness of the substrate are represented by $\rho_s$ and $l_s$, respectively, and those of the piezoelectric film are represented by $\rho_p$ and $l_p$. The measured $\Delta f_{overtone}$ is around 12.9 and 17.8 MHz for the HBAR on the Si (thickness 250 ± 5 μm) and SiC (thickness 350 ± 5 μm), respectively. After the acoustic velocity has been rectified using the aforementioned equation, the corrected acoustic velocities of the Si and SiC substrates are esteemed to be 6490 and 12500 m/sec, respectively. The acoustic velocity for the Si substrate is measured to be greater than the shear acoustic velocity value, despite the fact that it is remarkably equivalent to the reported longitudinal acoustic velocity value for the SiC substrate.[23,27] This multitude of modes offers a special opportunity to use the HBAR as a biofluid sensor.

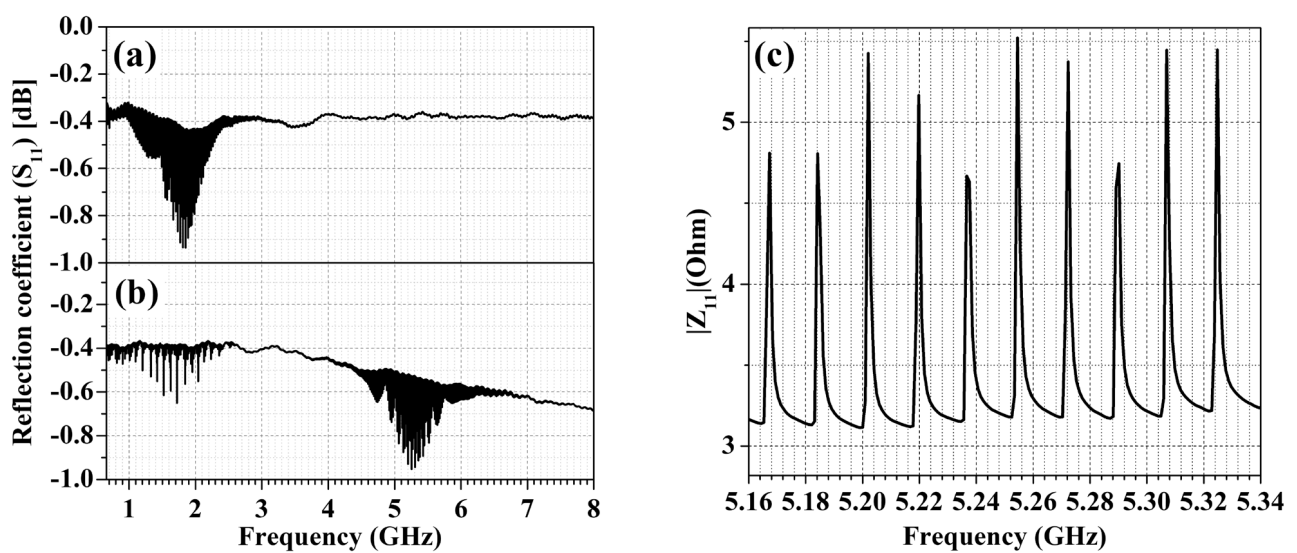

**Figure 3.** $S_{11}$ parameter for fabricated ZnO-based HBAR device on (**a**) Pt/Ti/*SiO$_2$*/Si and (**b**) Pt/Ti/SiC substrate. (**c**) $Z_{11}$ parameter for fabricated ZnO-based HBAR device on Pt/Ti/SiC substrate.

To comprehend the detailed behavior of device parameters, a modified Butterworth-Van Dyke (mBVD) model has been designed using Advance Design System (ADS, Keysight) software. Figure 4a depicts a typically measured and mBVD fitted $S_{11}$ parameter for ZnO-based HBAR on a SiC substrate at 5.25 GHz resonance. The mBVD model comprises circuit parameters such as motional resistance ($R_m$), motional capacitance ($C_m$), motional inductance ($L_m$), and static capacitance ($C_0$), resistance ($R_0$), which is shown as an equivalent circuit in Fig. 4b. From the equivalent circuit, the $f_r$ is the resonance frequency where the series resonance occurs, $f_a$ is the anti-resonance frequency where the parallel resonance occurs, and the effective electromechanical coupling coefficient ($k_{eff}^2$) is given by the below equations.

$$f_r = \frac{1}{2\pi\sqrt{L_m C_m}} \quad \text{and} \quad f_a = \frac{1}{2\pi\sqrt{L_m(C_m^{-1} + C_0^{-1})^{-1}}} \tag{2}$$

$$k_{eff}^2 = \frac{\pi^2}{4} \cdot \frac{f_r}{f_a} \cdot \left[1 - \frac{f_r}{f_a}\right] \tag{3}$$

The quality factor of the HBAR device on SiC substrate at the strongest excited resonance is measured using the new $Q$ approach based on $S_{11}$ parameter proposed by Feld et al. and it is related as follows;[28,29]

$$Q = \omega \cdot \left|\frac{dS_{11}}{d\omega}\right| \cdot \frac{1}{1 - |S_{11}|^2} \tag{4}$$

The electromechanical characteristics extracted from the mBVD model and the $Q$-factor of HBAR on Si and SiC substrates at the strongest excited resonances using the new $Q$ approach are listed in Table 1. The product of resonance frequency and the measured quality factor ($f.Q$ products) are then determined for the HBAR devices. We observed that the resonators for Si and SiC substrates, respectively, exhibit an $f.Q$ product of $0.06 \times 10^{13}$ and $4.1 \times 10^{13}$ Hz. Pang et al., Baumgartel et al. and Zhang et al. has reported that the $f.Q$ product of ZnO-based HBAR on Sapphire are $0.9 \times 10^{13}$, $4.5 \times 10^{13}$ and $4.8 \times 10^{13}$, respectively, using Lakin's $Q$ method (Table 2).[4,20,21] Here, we have also estimated the $f.Q$ value using Lakin's $Q$ method for HBAR devices on SiC, and reveals as $6.5 \times 10^{13}$ Hz, which, to the best of our knowledge stands out as the best among them. Using the novel $Q$ approach developed by Feld et al., the $f.Q$ product of ZnO-based HBAR on Diamond is reported by Gosavi et al. as $0.3 \times 10^{13}$, which is substantially lower than this finding[22].

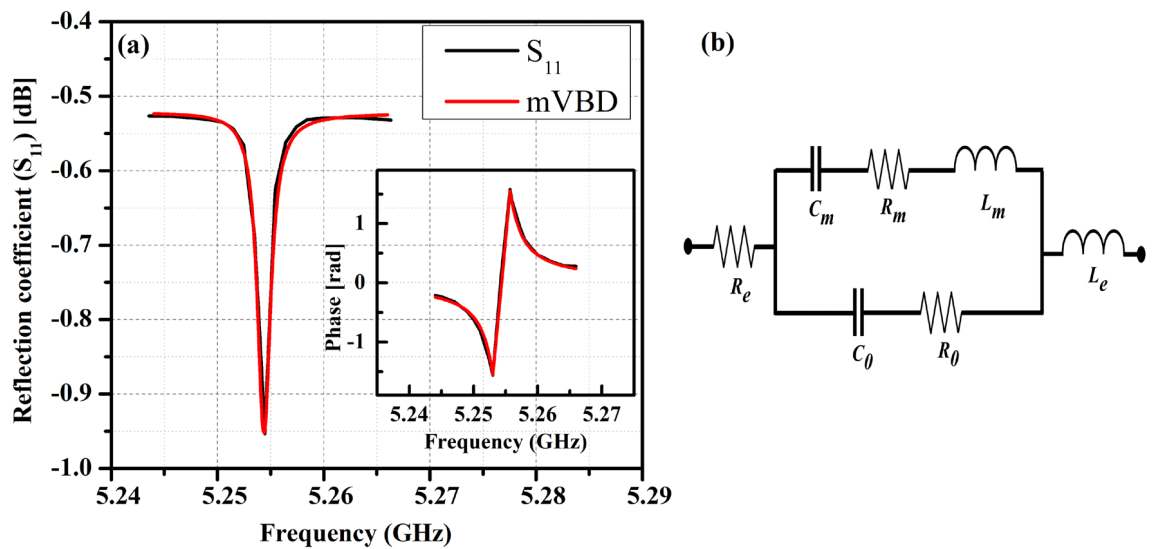

**Figure 4.** (**a**) The measured and mBVD fitted $S_{11}$ parameter and phase (inset) of ZnO-based HBAR resonator on SiC substrate at 5.25 GHz resonance, (**b**) The equivalent circuit diagram of mBVD model.

| ZnO-HBAR | $L_m$ (nH) | $C_m$ (pF) | $R_m$ (Ohm) | $C_0$ (pF) | $R_0$ (Ohm) | $R_e$ (Ohm) | $k_{eff}^2$ | $f$ (GHz) | $Q$ | $f.Q$ (Hz) |
|---|---|---|---|---|---|---|---|---|---|---|
| Si | 49.1 | 0.151 | 0.013 | 8.9 | 0.014 | 4.8 | 2.0 | 1.85 | 350 | $0.06 \times 10^{13}$ |
| SiC | 7.3 | 0.128 | 0.004 | 6.4 | 0.011 | 5.3 | 2.4 | 5.25 | 7725 | $4.1 \times 10^{13}$ |

**Table 1.** The measured and mBVD fitted $S_{11}$ parameters of ZnO-based HBAR resonator and their $f.Q$ values on Si and SiC substrate.

| HBAR | $f$ (GHz) | $f.Q$ (Hz) | References |
|---|---|---|---|
| Al/ZnO/Al on Sapphire | 2.43 | $0.9 \times 10^{13}$ (Lakin's Q method) | 20 |
| Mo/ZnO/Mo on Sapphire | 7.52 | $4.5 \times 10^{13}$ (Lakin's Q method) | 21 |
| Al/ZnO/Al on Sapphire, Quartz | 1 | 4.8 and $1.1 \times 10^{13}$ (Lakin's Q method) | 4 |
| Al/ZnO/Pt on Diamond | 1.45 | $0.3 \times 10^{13}$ (new Q approach) | 22 |
| Au/ZnO/Pt on SiC | 5.25 | $6.5 \times 10^{13}$ (Lakin's Q method) and $4.1 \times 10^{13}$ (new Q approach) | This work |

**Table 2.** ZnO-based HBAR resonator and their $f.Q$ values on various substrate.

## Conclusion

The acoustic properties at microwave frequencies have been investigated for ZnO-based HBARs on Si and SiC substrates. Highly *c*-axis oriented ZnO film is observed on Pt/Ti coated SiC substrate using the RF sputtering method. The novel *c*-axis oriented ZnO-HBAR on SiC substrate reveals multiple longitudinal bulk acoustic wave resonances and can be employed up to 7 GHz as a reliable wide-band piezoelectric transducer. The value of $f.Q$ product for HBAR on SiC substrate, determined using a unique Q approach method, is $4.1 \times 10^{13}$ Hz, which is superior to any reported $f.Q$ value among the specified ZnO-based HBAR on any other substrate. The findings will be valuable in the manufacture of both low-phase noise microwave oscillators and highly sensitive acoustic sensors.

## Methods and device fabrication

For the fabrication of HBAR, we have chosen a double-side polished, oxidized Si (100) and semi-insulating 4H-SiC (0001) wafer to achieve a metal insulator metal (MIM) capacitor configuration using an electron beam evaporator and dielectric magnetron sputtering equipment. For the bottom electrode of the HBAR, initially, 15 nm/100 nm thick Ti/Pt (adhesive/conducting) layers were deposited on cleaned wafers at room temperature using the electron beam evaporator technique. Then, a 650 ± 20 nm thick ZnO film was grown at 300 °C in an Ar:O$_2$ (1:1) gas atmosphere using an RF sputtering method. Wet etching of ZnO is accomplished using a photo-resist mask in order to shape the Pt bottom electrode. Finally, a stacked layer of Cr/Au with a thickness of 15/100 nm was deposited using an electron beam evaporator. This was followed by a lift-off photolithography procedure for the top electrode and active area patterning with 300 µm diameters. Figure 5 displays the final manufactured devices along with their material stack and microscopic picture. RF measurements on the developed HBAR were carried out using an Agilent vector network analyzer and a Ground–Signal–Ground probe station having an on-wafer pitch of 100 µm. On a typical standard substrate, calibration was carried out using the short, open, and load procedures. HBAR devices are measured as one-port devices in the frequency range of 0.5–10 GHz, with the top electrode functioning as the signal port and the bottom electrode functioning as the ground plane.

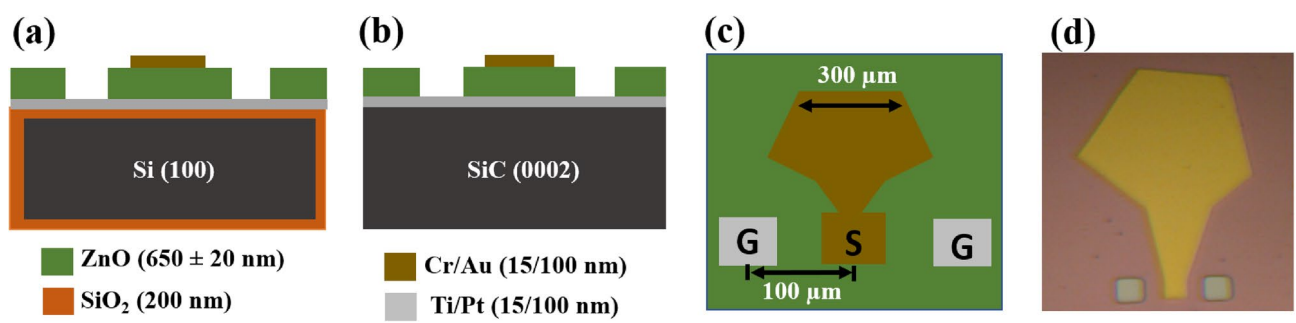

**Figure 5.** Device configurations and material stacks for HBARs fabricated on Si and SiC substrates (**a**) cross-sectional view of HBAR on Si, (**b**) cross-sectional view of HBAR on SiC, (**c**) top view of HBAR device, and (**d**) optical microscope image of the fabricated device.

## Acknowledgements
The authors would like to thank Ms. Akanksha Chouhan and Prof. Ashwin A. Tulapurkar for the Agilent vector network analyzer with a GSG probe station facility.

## Author contributions
P.P. carried out the experiments and wrote the main manuscript text. S.C. helped in measurements. S.T. and A.L. involve in conceptualizing and overall supervision of the works. All authors reviewed the manuscript.

## Competing interests
The authors declare no competing interests.

## Additional information
**Correspondence** and requests for materials should be addressed to A.L.

**Reprints and permissions information** is available at www.nature.com/reprints.